\documentclass[]{raa}            
\usepackage{graphicx,times}
\usepackage{natbib}

\providecommand{\bm}[1]{\mbox{\boldmath$#1$\unboldmath}}

\begin{document}

 \title{Observation of solar coronal heating powered by magneto-acoustic oscillations in a moss region}
\volnopage{ {\bf 2012} Vol.\ {\bf 9} No. {\bf XX}, 000--000}
\setcounter{page}{1}

\author{Parida Hashim\inst{1,2}, Zhenxiang Hong\inst{3}, Haisheng Ji\inst{3,4}, Jinhua, Shen\inst{2},  Kaifan Ji\inst{5}, Wenda, Cao\inst{6}}
\institute{Graduate School of the Chinese Academy of Sciences, Beijing, 100871, China\\
 \and Xinjiang Astronomical Observatory, CAS, Urumqi, 830011  \\
 \and Purple Mountain Observatory, DMSA/CAS, Nanjing, China, correspondence: jihs@pmo.ac.cn \\
 \and CfASS, China Three Gorges University, Yichang, 443002, China  \\
 \and Yunnan Astronomical Observatory, CAS, Kunming, 830011, China \\
 \and Big Bear Solar Observatory, 40386 North Shore Lane, Big Bear City, CA 922314, USA \\
            \vs \no
   {\small Received [2020] [xx] [xx]; accepted [xxxx] [xx] [xx]}
}

\abstract{In this paper, we report the observed temporal correlation between extreme-violet (EUV) emission and magneto-acoustic oscillations in a EUV moss region, which is the footpoint region only connected by magnetic loops with million-degree plasma. The result is obtained from a detailed multi-wavelength data analysis to the region with the purpose of resolving fine-scale mass and energy flows that come from the photosphere, pass through the chromosphere and finally heat solar transition region or the corona. The data set covers three atmospheric levels on the Sun, consisting of high-resolution broad-band imaging at TiO 7057 \AA\ and the line of sight magnetograms for the photosphere, high-resolution narrow-band images at Helium \textsc{i} 10830 \AA\ for the chromosphere and EUV images at 171 \AA\ for the corona.  The  10830 \AA\ narrow band images and the TiO 7057 \AA\ broad-band images are from a much earlier observation on July 22, 2012 with the 1.6 meter aperture Goode Solar Telescope (GST) at Big Bear Solar Observatory (BBSO) and the EUV 171 \AA\ images and the magnetograms are from observations made by Atmospheric Imaging Assembly (AIA) or Helioseismic and Magnetic Imager (HMI) onboard the Solar Dynamics Observatory (SDO).  We report following new phenomena: 1) Repeated injections of chromospheric material shown as 10830 \AA\ absorption are squirted out from inter-granular lanes with the period of $\sim$ 5 minutes.  2) EUV emissions are found to be periodically modulated with the similar periods of  $\sim$ 5 minutes.  3) Around the injection area where 10830 \AA\ absorption is enhanced, both EUV emissions and the strength of magnetic field are remarkably stronger. 4) The peaks on the time profile of the EUV emissions are found to be in sync with oscillatory peaks of the stronger magnetic field in the region. These findings may give a series of strong evidences supporting the scenario that coronal heating is powered by magneto-acoustic waves.
}
\authorrunning{Hashim et al.}
\titlerunning{Coronal heating powered by magneto-acoustic waves in a EUV moss region}
\maketitle

\section{Introduction}

A fundamental and long standing problem in astrophysics is why the tenuous solar (and many stellar) corona has a million-degree temperature far in excess of the underlying vastly denser photosphere. More and more evidence has revealed that the heating energy is generated directly from the photosphere via hot expulsions, spicules or Alfv$\acute{e}$n waves \citep{McIntosh-alfven, HTian2014, HPeter2014, Dinesh2019RAA}, but leaving no residual traces on it.  Resolution of this problem ultimately relies on resolving fine-scale mass and energy flows passing through the chromosphere, the interface layer sandwiched between the photosphere and the transition region or the corona \citep{iris, Tian2017RAA, DePontieu2009, Aschwanden2007}. 

By exploring correlations between the oscillatory signals observed at different levels of the solar atmosphere, many progresses have been made for resolving the mass and energy flows passing through the chromosphere, the interface layer \citep{iris, DePontieu2009, Aschwanden2007}. On the other hand, Helium \textsc{i} 10830 \AA\ narrow band imaging has been proven to be one of the best tools for precisely tracking mass and energy flows from below \citep{jicaogoode2012, Hong2017}. The Helium \textsc{i} 10830 \AA\ line is a weak absorption line seen against solar disk, yet the line is so special that the chromospheric line is associated with excitation by high-energy EUV photons. So, it is a line formed at the upper chromosphere. Since the line is shallow, a narrow band filtergram at this line taken near disk center contains the features of the photosphere like granules. Therefore, a big and crucial advantage for a narrow band filtergram taken at this line is that it can be precisely aligned with simultaneous images of the photosphere. 

Over the solar disk, the coronal heating rate differs on different parts of the solar surface, ranging from active regions to corona holes. An important target for studying the interface layer is footpoint regions of coronal loops, shown as faculae on the photosphere or plages in the chromosphere \citep{LiY2009RAA}. The region is believed to have a stronger heating rate, with prevalent oscillations at various wavebands \citep{DePontieu03}. In a plage area, of particular interest is the so-called EUV ``moss'', a region being connected to coronal loops with million-degree hot plasma.  In this region, much stronger heating rate is believed to be constantly occurring.  Nevertheless, the region has its name since it is full of dark inclusions from low temperature plasma making it take the appearance of reticulated bright EUV emission \citep{Berger1999, FletcherDePontieu1999}.  The dark inclusions are found to jointly appear and disappear, a signature of oscillatory fine-scale mass and energy flows going upward \citep{FletcherDePontieu1999, DePontieu03, DePontieu2006}. On the other hand, the oscillatory emissions of coronal loops as well as the rooted footpoint regions usually serve as indirect evidences of magneto-acoustic waves \citep{Jess2015, Banerjee2007, Nakariakov2005}.  The heating of the corona or plasma via damping of magneto-acoustic waves is of great interest across the entire community of astrophysics, as well as the plasma physics community \citep{Stix1975, Kolotkov2019}. However, as the first step, reliable correlations between the heating and magneto-acoustic oscillations have to be determined spatiotemporally with solid observations.  

On July 22, 2012, high-resolution narrow-band imaging at Helium \textsc{i} 10830 \AA\ for the chromosphere and broad band imaging at TiO 7057 \AA\ lines for the photosphere were carried out, targeting the active region NOAA 11259  (Ji Cao and Goode 2012, JCG12 hereafter).  The observations have revealed the smallest scale magnetic activities in inter-granular lanes of the photosphere with their brightening going up into the local transition region. The finding was confirmed by further analysis to the same data set or similar observations  (Zeng et al. 2013; Hong et al. 2017; Yang et al. 2019). In the most recent work, Ji et al. (2020) reported a possible correlation between the perturbations of magnetic field and He I 10830 \AA\ absorption, both have a quasi-periodic nature.  In the paper, they switch to concentrate on demonstrating that the magnetic field perturbations are magneto-acoustic oscillations powered by p-mode. In this paper, we give a thorough analysis to the correlation. We report the kind of peak-to-peak correlations between EUV emissions and magneto-acoustic oscillations. Observational results are given in section 2 and conclusions are given in section 3.

\section{Observational results}

 For details of the observation, we guide readers to the paper of JCG12. In this section, we will directly introduce the results analyzed from the observations. We have Fig. 1 here to show the observations made by different instruments. It includes the EUV moss region (the area of interest: AOI), its surrounding plages and the active region. The EUV moss region, with the area of  10 Mm $\times$ 10 Mm,  is inside the chartreuse-colored boxes in all panels (Fig. 1a-d). Fig. 1a and 1c show the EUV observation made by AIA at Fe IX 171 \AA\ on board SDO  and the X-ray telescope (XRT) on board Hinode taken with the Ti-poly filter \citep{Kosugi2007, GolubX}. In Fig. 1d we specially give the larger field of view to show the the morphology of the active region and its position on solar disk.   A moss region is usually situated at footpoints of SXR loops of million-degree plasma \citep{Berger1999, FletcherDePontieu1999}, so it is an ideal place for investigating the problem of coronal heating. A line-of-sight magnetogram observed by Helioseismic and Magnetic Imager (HMI) \citep{hmi-sdo} on board Solar Dynamic Observatory (SDO) \citep{SDO} is given to show the magnetic nature of the moss region (Fig. 1c). The magnetic field is of negative polarity, we will use unsigned magnetic field for the research in this paper. 
 
 \begin{figure}
\includegraphics[width=11cm]{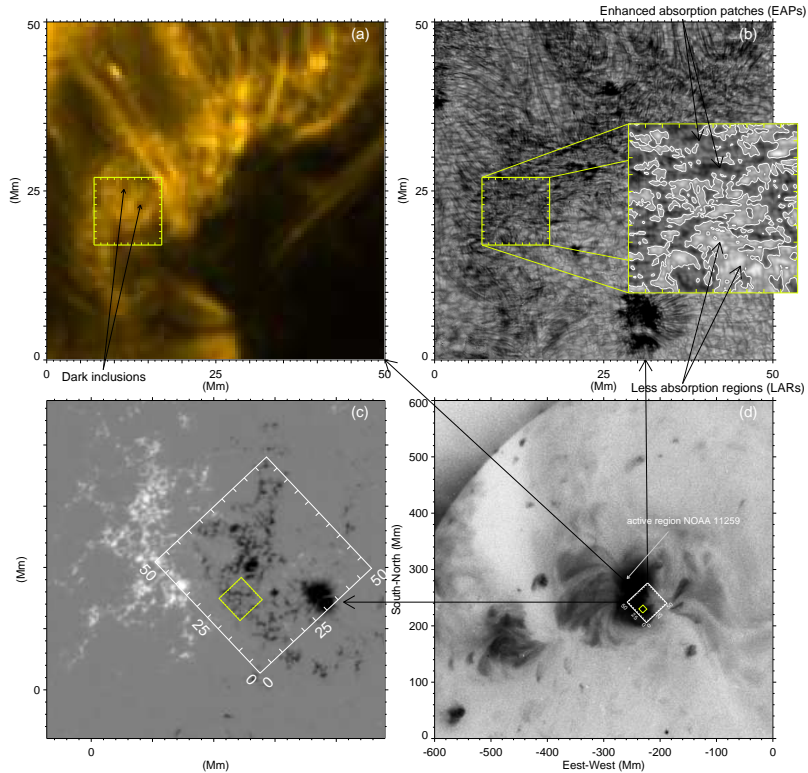}
\caption{\small  The chartreuse-colored boxes in all panels gives the site of the EUV moss region with the area of  10 Mm $ \times $ 10 Mm.  Panels a-c show the images at Fe IX 171 \AA\ observed by AIA, Helium \textsc{i} 10830 \AA\ observed by GST and a line-of-sight magnetogram observed by HMI, respectively.  In panel (b), an enlargement for the area of interest is specially given to show patches of enhanced 10830 \AA\ absorption and the patches with less 10830 \AA\ absorption (LAPs).   The EAPs are defined as the darkened areas within the contour level of $4.5 \times 10^3$ counts/pixel. Panel d, with a larger field view, gives a soft X-ray (SXR) image taken with the Ti-poly filter by the X-ray telescope (XRT) on board Hinode.}
\label{f1}
\end{figure}

 Panel (b) gives high-resolution imaging at narrow-band Helium \textsc{i} 10830 \AA\ for the region. The high-resolution reveals many small scale, and thus being ubiquitous, upward mass injections in the form of enhanced 10830 \AA\ absorption patches  (EAPs), as defined and discussed by Hong et al. (2017).  From panel a, we can see that point-like EUV dark inclusions from low temperature plasma are in the midst of EUV emissions. Dark inclusions are actually cool material being squirted upward as mass injections \citep{Berger1999, FletcherDePontieu1999}. As we will see in this paper, the dark inclusions, together with EUV emission, are spatiotemporally associated with enhanced Helium \textsc{i} 10830 \AA\ absorptions. The phenomenon signifies that the dark inclusions or the enhanced Helium \textsc{i} 10830 \AA\ absorptions are accompanied with some kind of heating processes.  In panel (b), an enlargement for the ROI is specially given to show the EAPs at 10830 \AA, which is within the contour level of $4.6 \times 10^3$ counts/pixel. We call other areas less 10830 \AA\ absorption  patches (LAPs). For the purpose of subsequent statistics, the value $4.6 \times 10^3$ counts/pixel is chosen in such a way that its contour level divides the ROI into two regions with roughly equal areas. The value corresponds to the $\sim$ 95\% of average band pass intensity for the moss region. 
 
As can be seen from the on-line animation in the paper of JCG12, there are numerous upward injections of Helium \textsc{i} 10830 \AA\ absorption material in the plage region. The injections actually repeat themselves in the same place periodically. The periodical nature can be demonstrated with a time-distance diagram made of a straight line across the area. Fig. 2-C2 gives a sample time-distance diagram at 10830 \AA\ from which we can see rows of periodical absorption structures. To get their spatial information on the photosphere, Fig. 2-A2 gives a time-distance diagram for the photosphere made of the vertical slice in the same place.  On the time-distance diagram, inter-granular lanes exhibit similar rows of parallel structures. The combination of the two kinds of time-distance diagrams made of Helium \textsc{i} 10830 \AA\ and TiO 7057 \AA\ images confirms that the mass injections are coming out of inter-granular lanes, or the boundaries of solar granules (Fig. 2-B2 or S2.mp4).  The associated EUV brightening shows the similar periodic structure for which we have just described. It is also worth noting that the rows of structures in the three time-distance diagrams (C2, B2 and D2) show similar moving patterns in the on-line animation. This phenomenon gives an additional proof demonstrating that the mass injections come out of inter-granular lanes. 

\begin{figure}
\includegraphics[width=11cm]{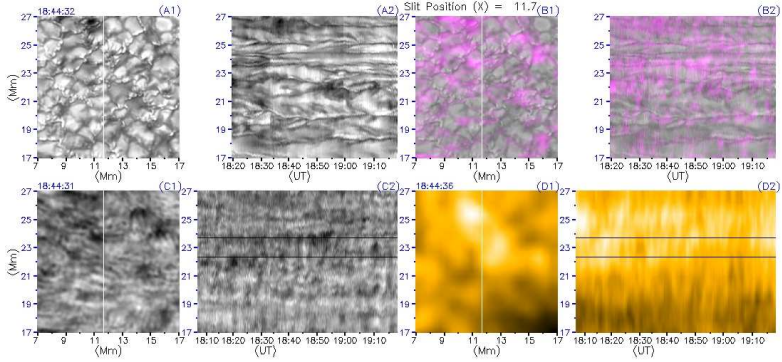}
\caption{\small  Panel A1 give the photosphere in the moss region as observed with a broad-band (10 \AA) filter at TiO 7057 \AA\ by GST (18:44:32 UT). Panels C1-D1 are the images of the moss region at Helium \textsc{i} 10830 \AA\ (18:44:31 UT) and Fe IX 171 \AA\ (18:44:36 UT) respectively.  Panel (B1) gives a composite image generated by taking the photospheric emission as background and putting 10830 \AA\ absorption (in pink) over it. Panels (A1-D1) have the same field view ($10 \times 10$ Mm$^2$) as the chartreuse-colored boxes in Fig. 1.  Panels (A2-D2) give 4 time-distance diagrams generated from the intensity distribution of the corresponding images along the vertical slit (the position of the vertical slit for this figure is at 11.4 Mm) shown in panels (A1-D1). The two sets of dark parallel lines are the boundaries of two slices being cut to become time profiles in Fig. 3a and c. On-line animation for varying time-distance diagrams generated with different slit positions is available.}
\end{figure}

In the same place sandwiched between the two dark parallel lines (Fig. 2-C2 and D2), we cut two slices out of the \textsc{i} 10830 \AA\ and EUV 171 \AA\ time-distance diagrams and covert them into two light curves. They are given in Fig. 3a and c. Wavelet analysis to the light curves give the oscillation period in the right two panels. They both have a component of $\sim$ 5 minutes, showing that they are somehow affected by the p-mode oscillation. We see that EUV emissions are correlated with the 10830 \AA\ absorption in a much complex manner.  The material injections from below, shown as EAPs, make contributions to dark inclusions. However, as we will see, they do heat the transition region to enhance EUV brightening at the same time. 

\begin{figure}
\includegraphics[width=11cm]{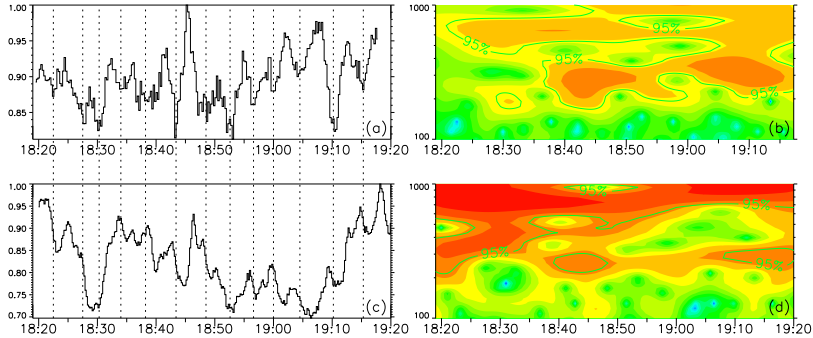}
\caption{\small Left two panels give two time profiles that are obtained from the rebinning of two slice time-distance diagrams shown in Fig. 2. The result of wavelet analysis to them is given in Right two panels. Upper two panels are for 10830 \AA\ while lower two panels are for 171 \AA. Vertical dashed lines indicate the times of the strongest 10830 \AA\ absorption.}
\end{figure}

Fig. 4a gives mean EUV emission at 171 \AA\ averaged over EAPs and LAPs in the moss region. We see that the EUV emission over the EAPs, shown with a red light curve, is systematically intenser over the region of EAPs. For emissions in other EUV wavelengths they have a similar behavior (Hong et al. 2017). We also compare strength of magnetic field in the EAPs and LAPs and the result gives that mean magnetic field is also systematically intenser over the region of EAPs (Fig. 4b). Furthermore, the two kinds of time profiles seems to have a kind of peak-to-peak correlation, when we compare the perturbations on the light curves. To exclude possibility that the temporal variations of mean magnetic field in Fig. 4b may be affected by the area of EAPs, which are also periodically changing, we computed another kind of time profile for the mean magnetic field for the region with the stronger field. For a certain time (say at 18:42:41 UT in this analysis), we select those pixels where the unsigned magnetic field is larger than 250 G. Then these pixels are collected together and get fixed to get the time profile in Fig. 4c. We see that the peak-to-peak correspondence still exists with the exception of a few EUV peaks.  With the help of a series of vertical lines in Fig. 14, we see that 14 out of the 18 EUV peaks can be regarded as being well correlated with the peaks of magnetic field perturbations. 

\begin{figure}
\includegraphics[width=11cm]{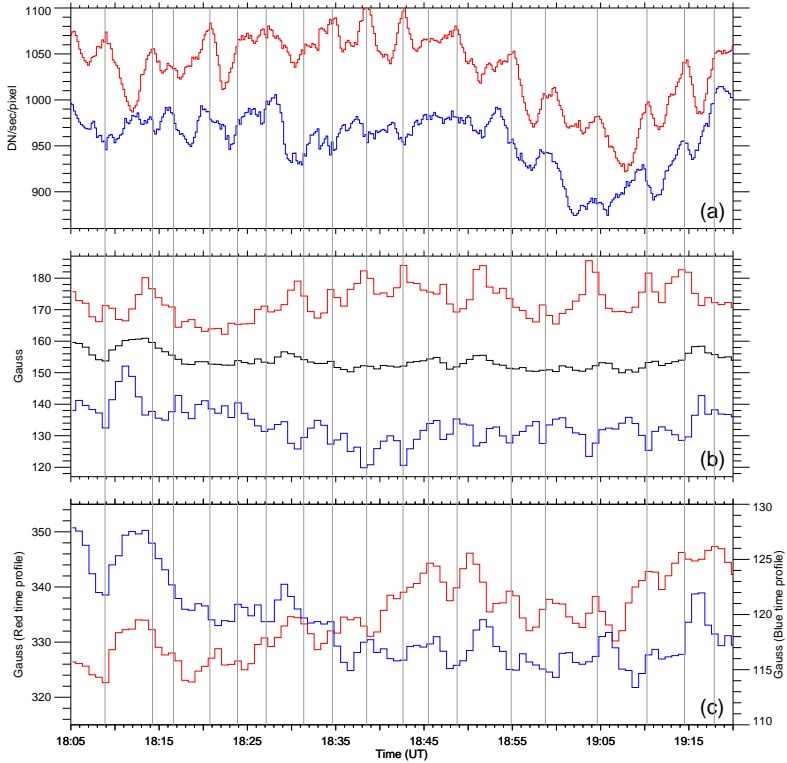}
\caption{\small Panels (a-b) give two kinds of time profiles for the mean EUV 171 \AA\ emission and mean magnetic field (unsigned) in the moss region, with red-colored light curve is over the EAPs and the blue is over the rest of regions (i.e. LAPs). In panel (b), the dark curve in the middle is the time profile of unsigned magnetic field averaged over the whole moss region. The red- and blue-colored light curves in the lowest panel are the time profiles for the mean unsigned magnetic field averaged over the stronger (larger than 250 G) and weaker  (lower than 250 G) magnetic field regions at 18:42:41 UT, respectively. Vertical grey lines run through peaks on the time profiles of the EUV emission over the EAPs}
\end{figure}

 \section{Discussion and Conclusions}

For high-resolution 10830 \AA\ narrow-band images, they are actually composite images containing dual features of the upper chromosphere and the photosphere.  There are numerous photospheric features which enable us to align the 10830 \AA\ narrow-band images with the images of the photosphere at TiO 7057 \AA\ with unprecedented precision. It can be called a natural co-alignment. Therefore, we can pin down the features of the chromosphere, 10830 \AA\ absorption in this paper, on the photosphere with least ambiguity. The precise co-alignment for high-resolution images has yielded the finding of magnetic activities in inter-granular lanes driven by granular convections (JCG12; Zeng et al. 2013; Hong et al. 2017). It is worth mentioning that, with high-resolution spectroscopic imaging in He I 10830 \AA\, Yang et al. (2019) reported that plasma flows along the magnetic loops are ejected from and drained out into inter-granule lane areas at different ends of magnetic loops.

With the same set of data as being observed and analyzed by JCG12, the paper aims to resolve fine-scale mass and energy flows in the chromosphere and their role in heating the corona. The mass and energy flows are signified by the periodic mass injections seen as 10830 \AA\ absorption in the selected moss region. For the periodic mass injections, they may also be the on-disk counterparts of spicules. For these periodic mass injections, we can conclude that

\begin{itemize}
  \item They have the period of $\sim$ 5 minutes,
  \item They are squirted out the area around the inter-granular lanes,
  \item They are accompanied by EUV brightening,
  \item They are associated with the stronger magnetic field.
\end{itemize}

In addition, the peaks on the time profile of the EUV emissions are found to be in sync with oscillatory peaks of the stronger magnetic field in the region. In our previous paper (Ji et al. 2020), we report that the magnetic perturbations are actually magneto-acoustic oscillations on the solar surface powered by the p-mode oscillation. For the two regions with the stronger and weaker magnetic field, the perturbations are frequently anti-phased. Therefore, signals of magnetic oscillations are usually reduced when the magnetic fluxes of the stronger field region and the weaker field region are integrated together. Only if we make a difference between the two kinds of regions can we get obvious signals of magnetic oscillations, and thus find their correlation with EUV emissions. 

All observations given in this paper give a series of proofs supporting that coronal heating in the plage region is powered by magneto-acoustic oscillations or waves. The magneto-acoustic waves may turn into upward propagating shocks when they meet a sharp change in density \citep{Hansteen, vanderVoort} or they may even modulate ongoing small-scale magnetic reconnection \citep{ChenPriest2006, Tian2019}. The existence of Alfven waves is worthy of further research using detailed phase analysis for these magnetic perturbations. The observations seems also support magnetic gradient pumping (MGP) mechanism \citep{Tan2014}. In the MGP mechanism, the magnetic gradient may drive the energetic particles to move upward from the underlying solar atmosphere and form hot upflows. To determine which mechanism really works for corona heating, current data is still insufficient in spatial resolution, temporal cadence and spectral coverage. With current large-aperture solar telescopes \citep{cao2010, bbsogst, LiuZhong},  spectro-polarimetry observations with higher resolution and high-cadence for the interface layer will be specially helpful to finally resolve the problem of coronal heating.

\begin{acknowledgements}
We thank the team SDO/AIA, SDO/HMI for providing the valuable data. SDO is a NASA project. The AIA and HMI data are downloaded via the Virtual Solar Observatory and the Joint Science Operations Center. We thank BBSO observing staff for their important assistance. This work is supported by NSFC grants 11333009, 11790302 (11790300), 11773061, 11729301, 11773072 and 11873027. 
\end{acknowledgements}

\newpage


\begin{thebibliography}{raa}
\bibitem[Aschwanden et al. 2007]{Aschwanden2007}Aschwanden, M.~J., Winebarger, A., Tsiklauri, D. et al., 2007, ApJ,  659, 1673
\bibitem[Berger et al. 1999]{Berger1999}Berger, T.~E., De Pontieu, B., Schrijver, C.~J. et al., 1999, ApJ,  519, L97
\bibitem[Banerjee et al. 2007]{Banerjee2007}Banerjee, D., Erd{\'e}lyi, R., \& Oliver, R., et al.\ 2007, \solphys, 246, 3
\bibitem[Cao et al. 2010]{cao2010}Cao, W., Gorceix, N., \& Coulter, R., et al.\ 2010, Astronomische Nachrichten, 331, 636
\bibitem[Chen \& Priest 2006]{ChenPriest2006}Chen, P.~F., \& Priest, E.~R.\, 2006, Solar Physics,  238, 313
\bibitem[De Pontieu \& Erd{\'e}lyi 2006]{DePontieu2006}De Pontieu, B., \& Erd{\'e}lyi, R.\ 2006, Phil. Trans. R. Soc. A, 364, 383
\bibitem[De Pontieu et al. 2003]{DePontieu03}De Pontieu, B., Erd{\'e}lyi, R., \& de Wijn, A.~G. 2003, ApJ,  595, L63
\bibitem[De Pontieu et al. 2009]{DePontieu2009}De Pontieu, B., McIntosh, S. W., Hansteen, V. H., \& Schrijver, C. J. 2009, ApJ,  701, L1
\bibitem[De Pontieu et al. 2014]{iris}De Pontieu B., Title, A. M., \& Lemen, J. R. et al.  2014, \solphys,  289, 2733
\bibitem[Dinesh Singh \& Singh Jatav 2019]{Dinesh2019RAA} Dinesh Singh, H. \& Singh Jatav, B.\ 2019, Research in Astronomy and Astrophysics, 19, 185
\bibitem[Fletcher et al. 1999]{FletcherDePontieu1999} Fletcher, L., \& De Pontieu, B., 1999, ApJ,  520, L135
\bibitem[Golub et al. 2007]{GolubX}Golub, L., Deluca, E., Austin, G. et al., 2007, Solar Phys.,  243, 63
\bibitem[Goode et al. 2010]{bbsogst}Goode, P.~R., Coulter, R., Gorceix, N., Yurchyshyn, V., \& Cao, W., 2010, Astron. Nachr.,  331, 620 
\bibitem[Hansteen et al. 2006]{Hansteen}Hansteen, V. H., De Pontieu, B., Rouppe van der Voort, L., van Noort, M., \& Carlsson, M. 2006, ApJ,  647, L73
\bibitem[Hong et al. 2017]{Hong2017}Hong, Z.-X., Yang, X., Wang, Y. et al., 2017, RAA,  17, 25
\bibitem[Jess et al. 2015]{Jess2015}Jess, D.~B., Morton, R.~J., Verth, G., et al.\ 2015, \ssr, 190, 103
\bibitem[Ji Cao and Goode 2012]{jicaogoode2012}Ji, H., Cao, W. \& Goode, P. R., 2012, ApJ,  750, L25 
\bibitem[Ji et al. 2020]{jietal2020}Ji, H. et al. 2020, ApJ, submitted
\bibitem[Kolotkov et al. 2019]{Kolotkov2019}Kolotkov, D.~Y., Nakariakov, V.~M., \& Zavershinskii, D.~I. 2019, A\&A,  628, A133
\bibitem[Kosugi et al. 2007]{Kosugi2007}Kosugi, T., Matsuzaki, K., Sakao, T., et al.\ 2007, \solphys, 243, 3
\bibitem[Lemen et al. 2012]{aia-sdo}Lemen, J.~R., Title, A.~M., \& Akin, D.~J. et al., 2012, Sol. Phys.,  275, 17  
\bibitem[Li and Ding 2009]{LiY2009RAA} Li, Y. \& Ding, M.-D.\ 2009, Research in Astronomy and Astrophysics, 9, 829
\bibitem[Liu et al. 2014]{LiuZhong} Liu, Z., Xu, J., Gu, B.-Z., et al.\ 2014, Research in Astronomy and Astrophysics, 14, 705-718
\bibitem[McIntosh et al. 2011]{McIntosh-alfven}McIntosh S. W. et al., 2011, Nature  475, 477
\bibitem[Nakariakov \& Verwichte 2005]{Nakariakov2005} Nakariakov, V.~M., \& Verwichte, E.\ 2005, Living Reviews in Solar Physics, 2, 3
\bibitem[Pesnell et al. 2012]{SDO}Pesnell, W.~D., Thompson, B.~J., \& Chamberlin, P.~C., 2012, Solar Phys.,  275, 3
\bibitem[Peter et al. 2014]{HPeter2014}Peter, H., Tian, H., Curdt, W. et al., 2014, Science,  346, 1255726
\bibitem[van der Voort et al.  2016]{vanderVoort}Rouppe van der Voort, L., De Pontieu, B., Pereira, T. M. D., Carlsson, M., \& Hansteen, V., 2016, ApJ,  799, L3
\bibitem[Samanta et al. 2019]{Tian2019}Samanta, T., Tian, H., \& Yurchyshyn, V. et al., 2019, Science,  366, 890
\bibitem[Schou et al. 2012]{hmi-sdo}Schou, J., Scherrer, P.~H., Bush, R.~I. et al., 2012, Sol. Phys.,  275, 327
\bibitem[Stix 1975]{Stix1975}Stix, T.~H., 1975, Nuclear Fusion,  15, 737
\bibitem[Tan 2014]{Tan2014} Tan, B.\ 2014, \apj, 795, 140
\bibitem[Tian et al. 2014]{HTian2014}Tian, H., DeLuca, E.~E., Cranmer, S.~R. et al., 2014, Science,  346, 1255711
\bibitem[Tian 2017]{Tian2017RAA} Tian, H.\ 2017, Research in Astronomy and Astrophysics, 17, 110
\bibitem[Yang et al. 2019]{2019ApJ...881L..25Y}Yang, X., Cao, W., Ji, H., et al.\ 2019, \apjl, 881, L25
\bibitem[Zeng et al. 2013]{Zeng2013}Zeng, Z., Cao, W., \& Ji, H.\ 2013, \apjl, 769, L33
\end{thebibliography}
\end{document}